\documentclass[12 pt]{article}

\usepackage{amsmath}
\usepackage{amsfonts}
\usepackage{amssymb}

\usepackage{graphicx}
\usepackage{epstopdf}

\usepackage{algorithm}
\usepackage{algpseudocode}

\usepackage{hyperref}
\usepackage[noabbrev,capitalise]{cleveref}
\crefformat{equation}{(#2#1#3)}

\DeclareMathOperator*{\argmax}{arg\,max}

\renewcommand\and{\end{tabular}\kern-\tabcolsep\ and\ \kern-\tabcolsep\begin{tabular}[t]{c}}
\let\origthanks\thanks
\renewcommand\thanks[1]{\begingroup\let\rlap\relax\origthanks{#1}\endgroup}

\author{Aurya Javeed\thanks{\scriptsize Center for Applied Mathematics, Cornell University, Ithaca, NY 14853, USA. (\href{mailto:gjh27@cornell.edu}{aj463@cornell.edu})} and Giles Hooker\thanks{\scriptsize Department of Statistical Science, Cornell University, Ithaca, NY 14853, USA. (\href{mailto:gjh27@cornell.edu}{gjh27@cornell.edu})}}

\title{Timing Observations of Diffusions}
\date{September 3, 2017}

\begin{document}
\maketitle

\begin{abstract}
  This paper addresses a problem in experimental design: We consider It\^o diffusions specified by some $\theta \in \mathbb{R}$ and assume that we are allowed to observe their sample paths only $n$ times before a terminal time $\tau < \infty$. We propose a policy for timing these observations to optimally estimate $\theta$. Our policy is adaptive (meaning it leverages earlier observations), and it maximizes the expected Fisher information for $\theta$ carried by the observations. In numerical studies, this design reduces the variation of estimated parameters by as much as 75\% relative to observations spaced uniformly in time. The policy depends on the value of the parameter being estimated, so we also discuss strategies for incorporating Bayesian priors over $\theta$.\\
  
\end{abstract}

\section{Introduction}
	
Suppose we have a process whose parameter we wish to estimate from a limited number of observations. We assume a model in the form of an It\^o diffusion,
\begin{equation}\label{eq:d}
  \mbox{d}\textbf{x}(t) = (\textbf{f} \circ \textbf{x}(t))\mbox{d}t + (\boldsymbol{\sigma} \circ \textbf{x}(t))\mbox{d}\textbf{w}(t),
\end{equation}
where $\textbf{w}(t)$ is Brownian motion of dimension $p$. Both $\boldsymbol{\sigma}$ and $\textbf{f}$ are deterministic, each being specified up to a real-valued constant $\theta$, which is our parameter of interest. We assume that sample paths of this system start from a known initial condition $\textbf{x}_0$. By extending \cite{hlr} and \cite{leifur}, we address the experimental design question: 
\begin{quote}
  Given only $n$ opportunities to observe \cref{eq:d} on the finite time interval $(0, \tau)$, what is the best way to budget the observations in real time so that we may estimate $\theta$?
\end{quote}
	
Our solution to this problem is described in \cref{sec:methods}. As in \cite{hlr} and \cite{leifur}, we compute an optimal policy using a dynamic program that maximizes the observations' expected $\theta$ information. Both references assume the diffusion is observed continually and focus on prescribing a $u_t$ that drives $\textbf{f}$, i.e., $\textbf{f} = \textbf{f}(\textbf{x}(t), u_t)$. In contrast, we assume that our system is costly to observe, limiting the number of sample path observations to $n$. Furthermore, we assume the diffusion evolves on its own accord (without an input), and that our observations can be chosen adaptively, meaning with knowledge of the previous observation times and outcomes.

For the purposes of simplifying the exposition, this paper only considers problems with one parameter of interest. If multiple parameters are to be estimated, the ideas we pursue extend directly to maximizing the trace of the expected information matrix. However, looking at more than one parameter makes it difficult to assess the effectiveness of the design because we might trade accuracy in one parameter in favor of another. Other, non-linear, functions of the Fisher information could be approximately examined, but this confuses our development of dynamic programing with the selection of an objective function.

As concrete examples, this paper considers two systems for which observations are limited: (a) the concentration of a drug in a patient's bloodstream and (b) the number of algae and rotifers trapped in a chemostat. The former is measured with blood samples, while the latter requires a trained scientist to count organisms under a microscope. We study these systems in \cref{sec:examples}, in addition to a process with slow and fast timescales.
	
We will see that our optimal policy depends on the very parameter, $\theta$, that it is designed to estimate. \cref{sec:dependence} addresses this limitation by adapting the Bayesian machinery proposed by \cite{hlr} and \cite{leifur}. \cref{sec:conclusion} concludes.
\section{Methods}\label{sec:methods}

The observation times that we consider optimal are those that maximize the expected Fisher information about $\theta$ (our parameter of interest). Since we will be estimating $\theta$ with maximum likelihood (ML) estimates, this definition of optimality is equivalent to minimal estimator variance when the number of observations $n \to \infty$. This definition is also reasonable for small $n$: the expected Fisher information is the mean curvature of the likelihood, hence the larger it is, the more pronounced the $\theta$ that best fits the data (on average).

The design we choose is adaptive. In dynamical systems, adaptivity is important in order to improve our forecast of the state of the system at the next observation time. Thus, for the $i$th observation, we produce a policy, $\hat{t}_i(s, \textbf{x})$, that returns the next observation time, $\hat{t}_i$, based on the current observation being taken at time $s$ and finding the state of the system at $\textbf{x}$. This policy is introduced in \cref{sec:des}; conveniently it can be precomputed, stored, and implemented in real time. \cref{sec:imp} provides numerical details.

\subsection{Proposed Design}\label{sec:des}

Since \cref{eq:d} is Markov, we can unravel our optimization problem with a \emph{dynamic program}, building the optimum backward from the last observation to the first. We begin with the $\theta$ Fisher information carried by a single observation:
\begin{align}\label{eq:i}
  \mathcal{I}(t, \textbf{x}_0) = \mathbb{E}_{\textbf{y} | (t, \textbf{x}_0)} \left[ \bigg( \frac{\partial}{\partial\theta} p(\textbf{y}, t | \textbf{x}_0) \bigg)^2 \right],
\end{align}
where
\begin{align*}
\mathbb{E}_{\textbf{y} | (t, \textbf{x}_0)} [\,\cdot\, ] := \int (\,\cdot\,) p(\textbf{y}, t | \textbf{x}_0)d\textbf{y}.
\end{align*}
In these expressions, $p$ is the transition density of \cref{eq:d}, such that $p(\textbf{y}, t | \textbf{x}_0)$ denotes the time-$t$ density of $\textbf{y}$ after initializing at $\textbf{x}_0$. Notice that the expected information, \cref{eq:i}, does not depend on the outcome of the observation, only the time it occurs and the initial condition of the process.

Because \cref{eq:d} is time-homogeneous, we can generalize \cref{eq:i} to a starting state $\textbf{x}$ at a time $s < t$ (with $s$ not necessarily equal to zero). In this setting, \cref{eq:i} becomes
\begin{align*}
  \mathcal{I}(t - s, \textbf{x}).
\end{align*}
Thus, from $(s, \textbf{x})$ onward, the maximal information an observation can attain before a time $\tau$ is
\begin{align*}
  \mathcal{M}_1(s, \textbf{x}) = \sup_{t \in (s, \tau)} \mathcal{I}(t - s, \textbf{x}).
\end{align*}
The most information two observations can carry is
\begin{align*}
  \mathcal{M}_2(s, \textbf{x}) = \sup_{t \in (s, \tau)} \left\{ \mathcal{I}(t - s, \textbf{x}) + \mathbb{E}_{\textbf{y} | (t, \textbf{x})} \mathcal{M}_1(t, \textbf{y}) \right\},
\end{align*}
and continuing recursively, we obtain
\begin{align}\label{eq:m}
  \mathcal{M}_i(s, \textbf{x}) = \sup_{t \in (s, \tau)} \left\{ \mathcal{I}(t - s, \textbf{x}) + \mathbb{E}_{\textbf{y} | (t, \textbf{x})} \mathcal{M}_{i - 1}(t, \textbf{y}) \right\}, 
  \quad \mathcal{M}_0 := 0.
\end{align}
In this equation, the term $\mathcal{I}(t - s, \textbf{x})$ is the information carried by the first observation after $(s, \textbf{x})$. The second term, $\mathbb{E}_{\textbf{y} | (t, \textbf{x})} \mathcal{M}_{i - 1}(t, \textbf{y})$, is the maximal information expected thereafter. Consequently we say that $\mathcal{M}_i$ is the maximal \emph{Fisher information to go} (FITG) for $i$ observations, a phrase adopted from \cite{hlr} and \cite{leifur}.\footnote{We take the expectation of the maximal FITG (rather than the maximum of the expected FITG) since observations are allowed to be chosen adaptively; that is, once we know $\textbf{y}$, we are free to go after the largest possible information that remains.}

Without loss of generality, we assume that, for all $s$ and $i$, the supremum in \cref{eq:m} is achieved on the interval $(s, \tau)$. Therefore, the optimal observation times are
\begin{align}\label{eq:t}
  \hat{t}_i(s, \textbf{x}) = \argmax_{t \in (s, \tau)} \left\{ \mathcal{I}(t - s, \textbf{x}) + \mathbb{E}_{\textbf{y} | (t, \textbf{x})} \mathcal{M}_{n - (i - 1)}(t, \textbf{y}) \right\}.
\end{align}
Here the subscript of $\mathcal{M}$ is reindexed so that $i = 1$ specifies the first observation and $i = n$ the last. The optimal times vary with their predecessors (adaptivity), but there is no explicit dependence on earlier ancestors (Markov property).

Our proposed policy follows as \cref{alg:pol}. In the event that its maximum is not unique, we take the smallest maximizer.
\begin{algorithm}
\caption{Optimal Observation Times}\label{alg:pol}
\begin{algorithmic}[1]
\State{Set $t_0 = 0$.}
\For{$k = 1$ through $n$}
  \State Set $t_k = \hat{t}_k(t_{k - 1}, \textbf{x}_{k - 1})$ as the $k$th optimal observation time.
  \State Observe the sample path of \cref{eq:d} at time $t_k$, and store the result as $\textbf{x}_k$.
\EndFor
\end{algorithmic}
\end{algorithm}

\subsection{Numerical Implementation}\label{sec:imp}

To compute our policy, we need to know the transition density, $p$, which appears explicitly in the definition of $\mathcal{I}$ and implicitly in the expectations throughout the previous subsection. Generally the density is not available analytically, so we approximate it numerically by discretizing \cref{eq:d} over a finite state space $S$ using a \emph{locally consistent} Markov chain (i.e., a chain whose steps, to first order in time, have the same mean and covariance as increments of the diffusion).

We build the chain according to \cite{kushner}. The construction requires that
\begin{enumerate}
 \item[(i)] $S$ be a rectangular lattice whose vertices are separated by a uniform distance $h$, and 
 
 \item[(ii)] $\boldsymbol{\sigma}(\textbf{x})\boldsymbol{\sigma}(\textbf{x})^T$ be diagonal. 
\end{enumerate}
Should these assumptions be too restrictive for a given application, more general derivations are also described in \cite{kushner}.

The linchpin of our construction is the \emph{Kolmogorov backward equation} governing the transition density, $p$:
\begin{align}\label{eq:b}
	-\partial_t p(\textbf{x}, t) = \textbf{f}(x) \cdot \nabla p(\textbf{x}, t) 
	+ \frac{1}{2} (\boldsymbol{\sigma}(\textbf{x}) \boldsymbol{\sigma}(\textbf{x})^T) : \nabla^2 p(\textbf{x}, t).
\end{align}
This partial differential equation is discretized with backward differences for $\partial_t p$, first-order upwind differences for $\nabla p$, and centered differences for elements of $\nabla^2 p$. Letting $\delta$ and $h$ denote the time and space increments, the discretization of \cref{eq:b} at $(t + \delta, \textbf{x})$ is equivalent to
\begin{align}\label{eq:disc}
	\begin{split}
		p(\textbf{x}, t) &= p(\textbf{x}, t + \delta) \left[ 1 - \text{tr}(\boldsymbol{\sigma}(\textbf{x})\boldsymbol{\sigma}(\textbf{x})^T) \frac{\delta}{h^2} - \sum_{i = 1}^d |f_i(\textbf{x})| \frac{\delta}{h} \right] \\
		&+ \sum_{i = 1}^d \Bigg\{ p(\textbf{x} + h\textbf{e}_i, t + \delta) \left[ \frac{(\boldsymbol{\sigma}(\textbf{x})\boldsymbol{\sigma}(\textbf{x})^T)_{ii}}{2} \frac{\delta}{h^2}+ f_i^+(\textbf{x}) \frac{\delta}{h} \right] \hspace{0 in} \\
		&\hspace{0.75 in} + p(\textbf{x} - h\textbf{e}_i, t + \delta) \left[ \frac{(\boldsymbol{\sigma}(\textbf{x})\boldsymbol{\sigma}(\textbf{x})^T)_{ii}}{2} \frac{\delta}{h^2} + f_i^-(\textbf{x})\frac{\delta}{h} \right] \Bigg\}.
	\end{split}
\end{align}
Here $f_i$ is the $i$th component of $\textbf{f}$, $f_i^{\pm} = \max(\pm f_i, 0)$, and $\textbf{e}_i$ is the $i$th vector in the standard basis of $\mathbb{R}^d$, with $d := \dim(\textbf{x}(t))$. When $\delta$ is small enough, all coefficients of $p$ on the right-hand side of \cref{eq:disc} are positive. Since they also sum to one, these coefficients are interpreted as the transition probabilities of a Markov chain. For instance,
\begin{align*}
	\frac{(\boldsymbol{\sigma}(\textbf{x})\boldsymbol{\sigma}(\textbf{x})^T)_{ii}}{2} \frac{\delta}{h^2} + f_i^-(\textbf{x})\frac{\delta}{h}
\end{align*}
is the coefficient of $p(\textbf{x} - h\textbf{e}_i, t + \delta t)$, and thus is the probability of moving from $\textbf{x}$ to $\textbf{x} - h\textbf{e}_i$ after a step $\delta$. Similarly, the probability of staying in place after increment $\delta$ is
\begin{align*}
  1 - \text{tr}(\boldsymbol{\sigma}(\textbf{x})\boldsymbol{\sigma}(\textbf{x})^T) \frac{\delta}{h^2} - \sum_{i = 1}^d |f_i(\textbf{x})| \frac{\delta}{h}.
\end{align*}

Verifying that the resulting chain is locally consistent with \cref{eq:d} is straightforward. However, the set of states, $S$, must have finite bounds. To manage this, we use the same convention as \cite{hlr}, wherein the chain is forced to stay in place if it tries to move off $S$ (i.e., we fold the probabilities of leaving the grid into those of not moving). This truncation is not significant, as $S$ can often be chosen large enough that the probability of exit is small.

The chain's transition matrix, $P$, is used to approximate the diffusion's transition density, $p$; namely, for any two $\textbf{s}_i$ and $\textbf{s}_j \in S$, $p(\textbf{s}_j, \delta k | \textbf{s}_i)$ is taken to be $[P^k]_{ij}$.

\subsubsection{Additional Details}
\paragraph{Discretizing Time:} Often the $\delta$ needed to make the coefficients in $\cref{eq:disc}$ positive is so small that our policy scarcely changes across time increments of that size. For greater computational efficiency, we discretize time with the mesh
\begin{align*}
	T = 0:\gamma\delta:(\tau - \gamma\delta),
\end{align*}
where $\gamma$ is a dilation factor that divides $\tau\delta^{-1}$. 

\paragraph{Sample Paths:} To simulate sample paths of the diffusions we observe, we use an Euler-Maruyama integrator with steps of size $\gamma\delta/10$. Observations of these sample paths generally do not belong to $S$ (the grid of states on which the optimal observation times are defined), but we overcome this by rounding observations to the closest element of $S$.

\paragraph{Parameter Estimates:} After collecting our set of observations, we approximate $\theta$ with an ML estimate on a grid of candidate values, $\Phi$. At a candidate $\phi \in \Phi$, we build the log-likelihood from the powers of transition matrix $P$ with $\theta$ set equal to $\phi$. The run time of this construction is bottlenecked by multiplying $P$ with itself.

\section{Examples}\label{sec:examples}

We demonstrate our policy with three different diffusions. The first two model real-world systems that are costly to observe, while the third is a mathematically clean example that further elucidates our proposed policy.
  
\subsection{Pharmacokinetics}

The first diffusion we study is from the field of \emph{pharmacokinetics}, which studies the movement of drugs in organisms. A \emph{single compartment} model reduces an organism to a single unit and assumes that a one-time drug dose is absorbed into the blood stream at a rate roughly proportional to its unabosrbed concentration, i.e.,
\begin{align}\label{eq:sc}
  \dot{x} = -\alpha x.
\end{align}
We perturb \cref{eq:sc} by a Brownian increment to emulate model misspecification and the stochastic fluctuations observed in empirical data. The result is
\begin{align}\label{eq:ou}
  dx = -\alpha x dt + \sigma dw_t,
\end{align}
with $\alpha$ our parameter of interest. Mathematically this system is an \emph{Ornstein-Uhlenbeck} process, one of the simplest instantiations of \cref{eq:d}. 

To prescribe reasonable parameter values for \cref{eq:ou}, we used the R data set \texttt{Theoph} \cite{r}. It gives the concentrations of the anti-asthmatic drug \emph{theophylline} in 12 subjects, over the course of 25 hours after they were administered a one-time dose. We fit \cref{eq:sc} to Subject 11, and find that $\alpha = 2$ with an initial condition of $x_0 = 8$. We set the noise amplitude $\sigma = 1$ and allow ourselves $n = 3$ observations until a time horizon of $\tau = 2$ days (which is approximately twice the duration of the \texttt{Theoph} data set).

\begin{figure}
  \centering
  \includegraphics[width = \textwidth]{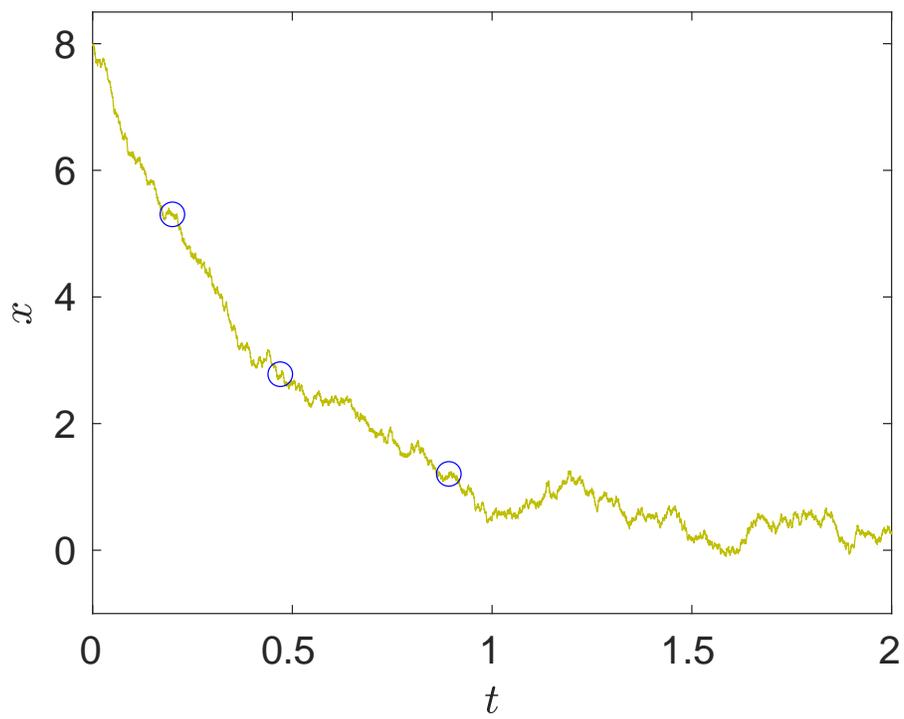}
  \caption{A sample path of the pharmacokinetic model \cref{eq:ou}. The initial condition $x_0 = 8$, and the $n = 3$ observations chosen by our policy are circled in blue.}\label{fig:pkm}
\end{figure}

\cref{fig:pkm} shows the three observations that our policy selects on a generic actualization of \cref{eq:ou}.\footnote{For the discretization, we set $S = \texttt{-2:0.01:9}$, $\delta = 10^{-5}$, and $\gamma = 200$.}
Recall that the entire sample path is not available to the policy; it knows only the states and times of previous observations, beginning with Observation 0 $:= x_0$.

Before going further, we make two remarks about the behavior we expect of an optimal policy:
\begin{enumerate}
  \item Notice that $|\alpha x| \gg |\sigma dw_t|$ when $x$ is large. Thus observations at $|x| \gg 0$ should convey the most information about $\alpha$.

  \item This said, if observation times are not sufficiently separated from one another, the noise term will obfuscate the $-\alpha x$ decay whose rate we aim to estimate. 
\end{enumerate}
Therefore a good policy should choose observations far from zero without stacking them immediately after each other.

Because the state space of \cref{eq:ou} is one-dimensional, we can visualize each of our optimal observation times, $\hat{t}_i$, with a heat map.
\begin{figure}
  \centering
  \includegraphics[width = 0.327\textwidth]{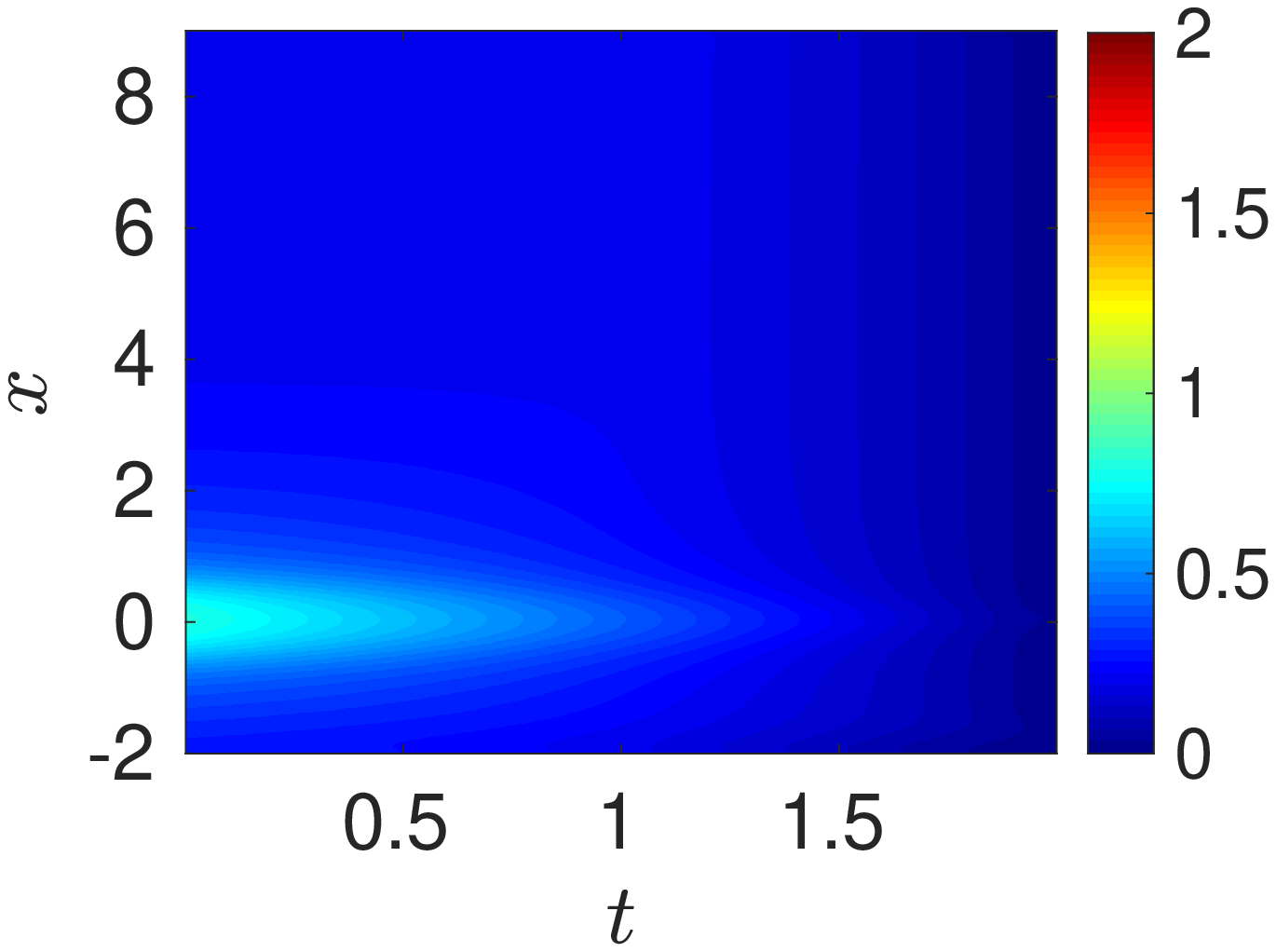}
  \includegraphics[width = 0.327\textwidth]{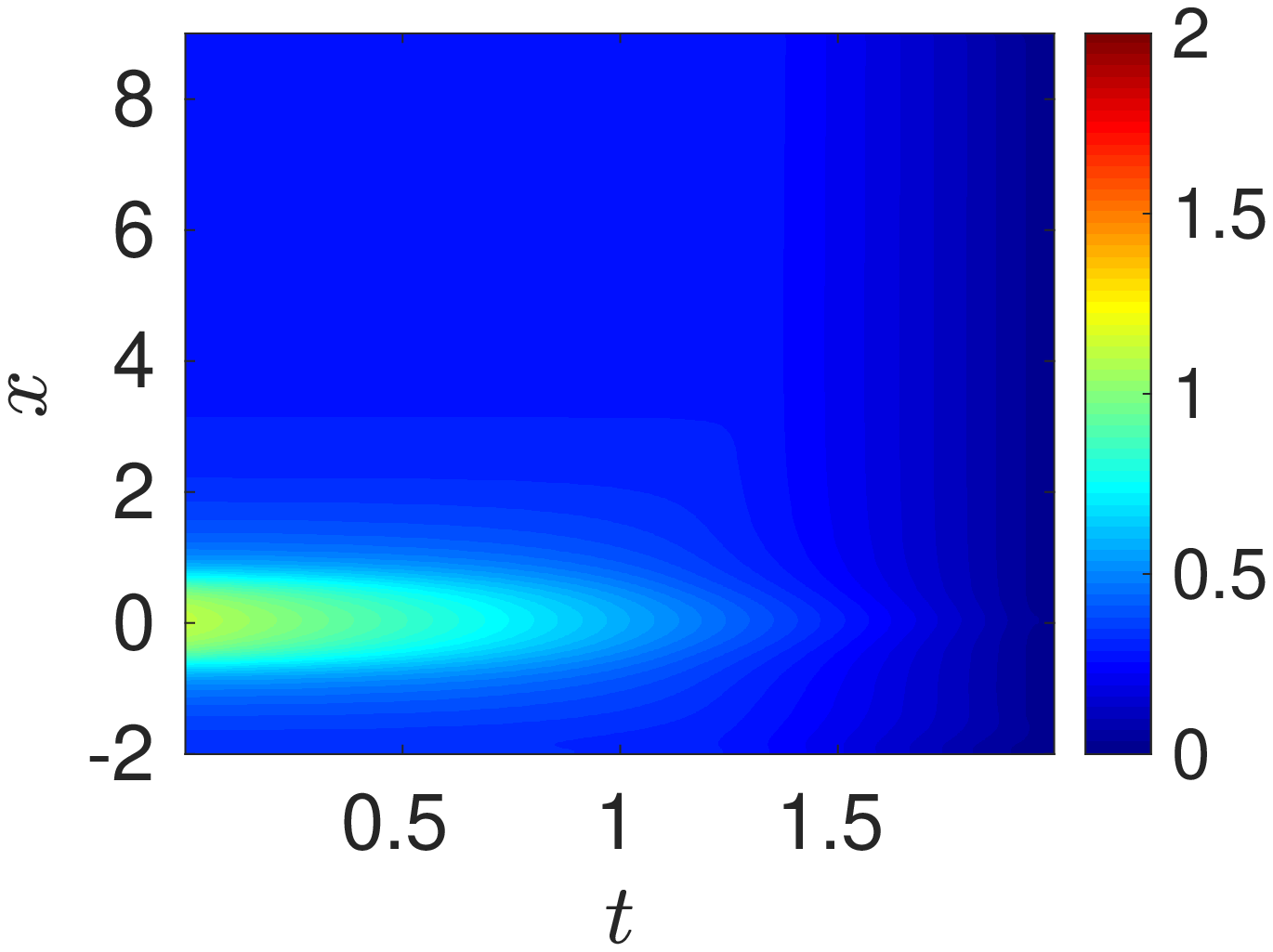}
  \includegraphics[width = 0.327\textwidth]{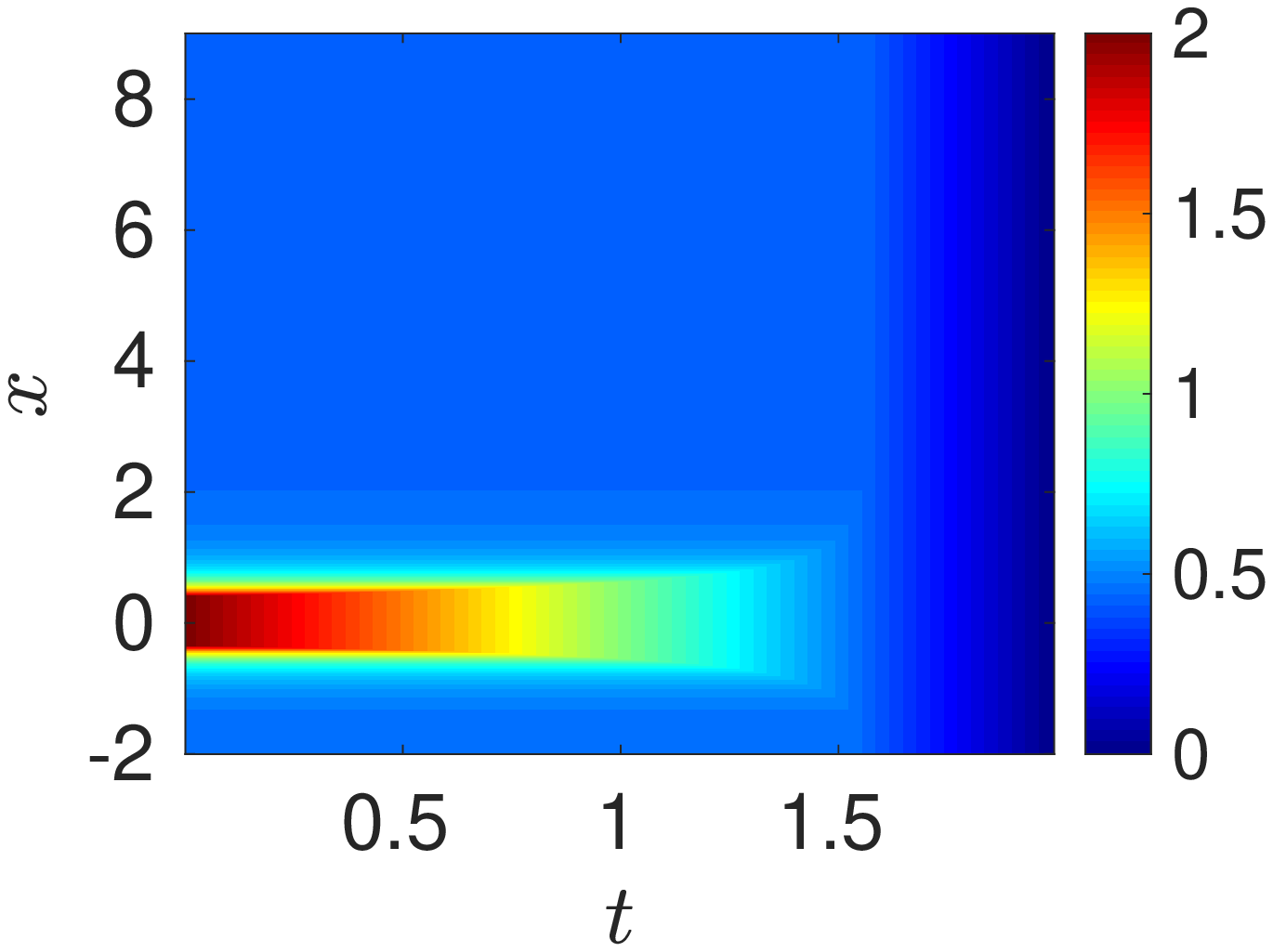}
  \caption{The optimal $\hat{t}_i(t, x)$ for observing \cref{eq:ou}. From left to right, $i = 1$, $2$, and $3$.}\label{fig:pol}
\end{figure}
\cref{fig:pol} reveals that our optimal design does indeed abide by the two expectations above. In particular, the sea of blue implies Observation $i$ is taken quickly when Observation $i-1$ finds the process far from zero; however the two observations are not stacked in short succession unless there is not much time until $\tau$. The figure also shows that an observation is delayed when its predecessor is close to zero. This is reasonable since \cref{eq:ou} is virtually stationary near zero, ergo we do not expect any observation time to be better than any other.

We construct ML estimates of $\alpha$ on the grid $\Phi$ spanning 0.1 to 10 in increments of 0.1. For the single realization of our process and policy shown in \cref{fig:pkm}, we find $\hat{\alpha} = 2.2$. Reapplying our policy to 99 additional sample paths, we obtain the sample statistics presented in row 1 of \cref{tbl:pkm}. As a point of comparison, row 2 of \cref{tbl:pkm} gives the sample statistics when the same paths are observed at the three times spaced uniformly across $(0, \tau)$.
\begin{table}
  \caption{Sample statistics of $\hat{\alpha}$ across 100 actualizations with $n = 3$. In parentheses, we include the statistics when the process is initialized close to stationarity---i.e., when $x_0 = 0$.}\label{tbl:pkm}
  \centering
  \begin{tabular}{ c || c c | c c }
     & \multicolumn{2}{c |}{Bias} & \multicolumn{2}{c}{Standard Deviation} \\ \hline\hline
    Policy           & 0.057 & (2.011) & 0.264 & (3.322) \\
    Uniformly Spaced & 0.102 & (1.720) & 0.358 & (3.093) \\
  \end{tabular}
\end{table}
The bias (respectively, variance) of our policy's $\hat{\alpha}$ is 44\% (26\%) smaller than in the case of uniform observations. The mean-square error (MSE) of $\hat{\alpha}$ is
\begin{align*}
  \mathbb{E}[(\hat{\alpha} - \alpha)^2] = \text{var}(\hat{\alpha}) + \text{bias}(\hat{\alpha}, \alpha)^2,
\end{align*}
which equals 0.073 and 0.139 for the two techniques.

\begin{figure}
  \centering
  \includegraphics[width = 0.49\textwidth]{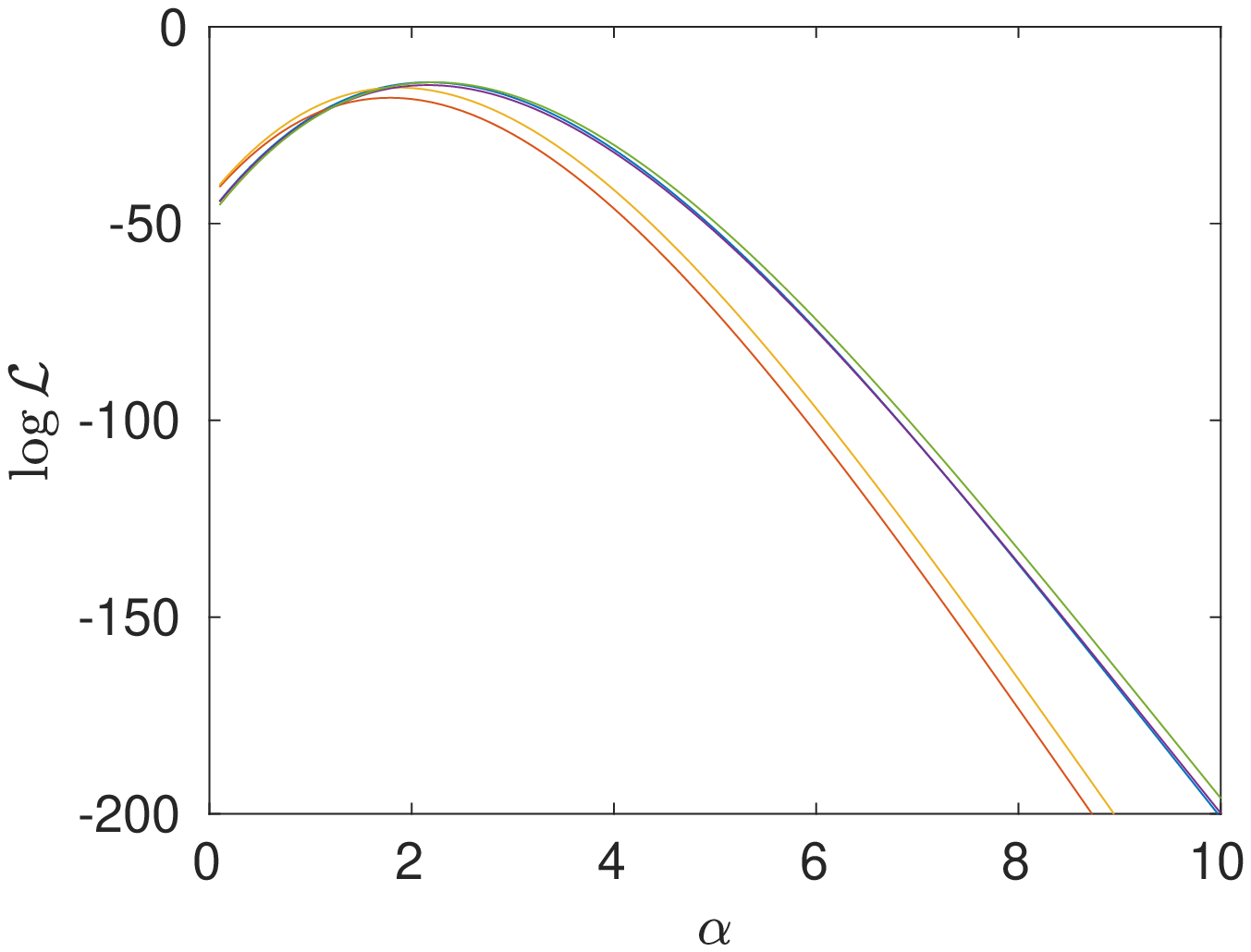}
  \includegraphics[width = 0.49\textwidth]{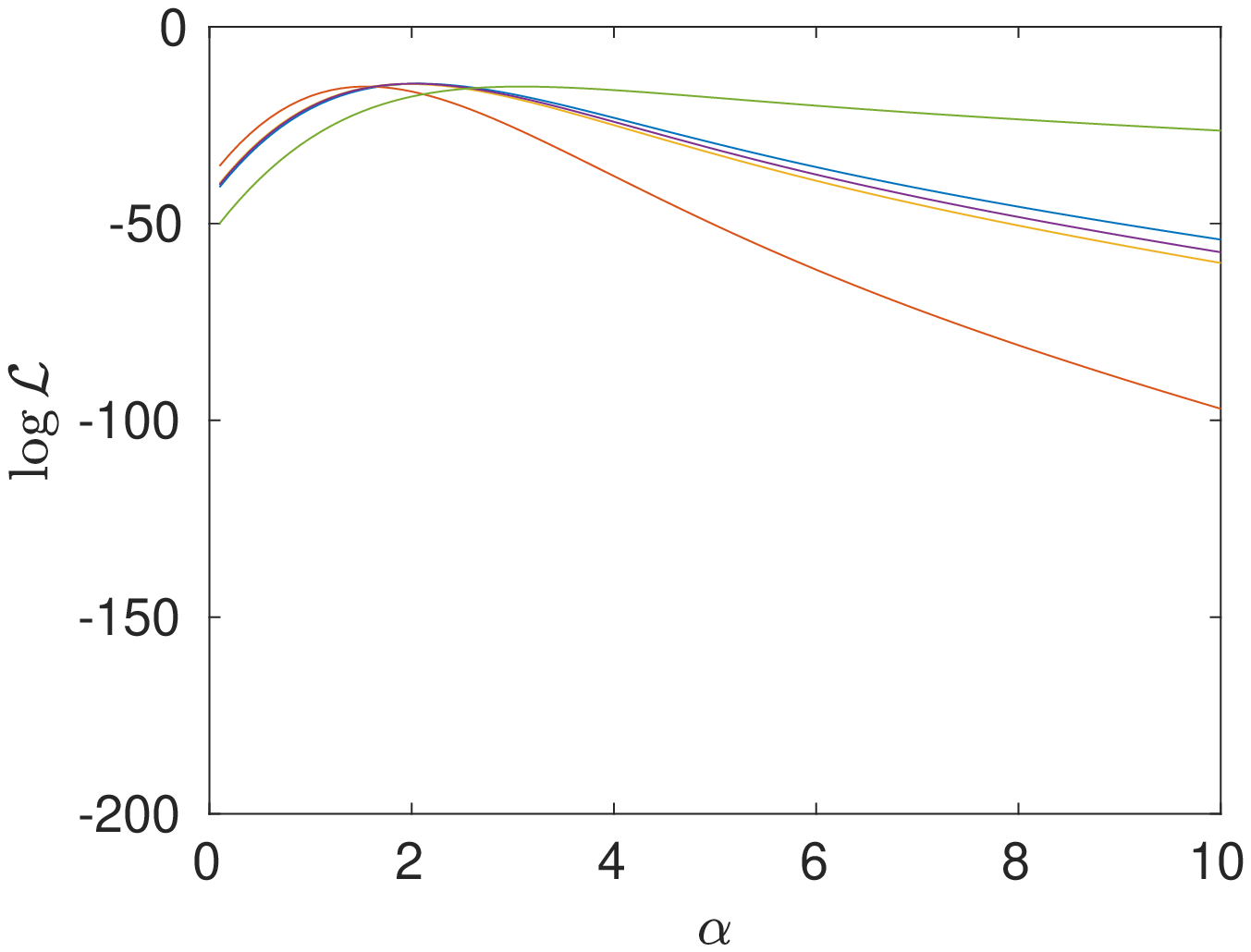}
  \caption{Pharmacokinetic log-likelihoods under our policy (left) and uniformly-spaced observations (right), for the same five randomly selected sample paths. $n = 3$.}\label{fig:lik}
\end{figure}

At this point, two comments are in order:
\begin{enumerate}
  \item As suggested earlier, we do not expect our policy to provide a substantial advantage when it is applied to a process close to stationarity. The parentheticals of \cref{tbl:pkm} give the statistics when $x_0$ is changed to zero---the mode of the process's stationary distribution.
  
  \item \cref{fig:n} provides a sense of how our policy fares for different $n$. The policy is guaranteed to minimize the asymptotic variance of $\hat{\alpha}$, but \cref{fig:n} shows that the improvement is small compared to taking observations spaced uniformly in time. The biggest gains come at small $n$. Thus, as we claimed in \cref{sec:des}, our policy is reasonable for limited observations, despite offering no guarantee of minimal estimator variance. Compared to uniformly-spaced observations, \cref{fig:lik} shows that the policy sharpens the likelihood substantially.
\begin{figure}
  \centering
  \includegraphics[width = 0.49\textwidth]{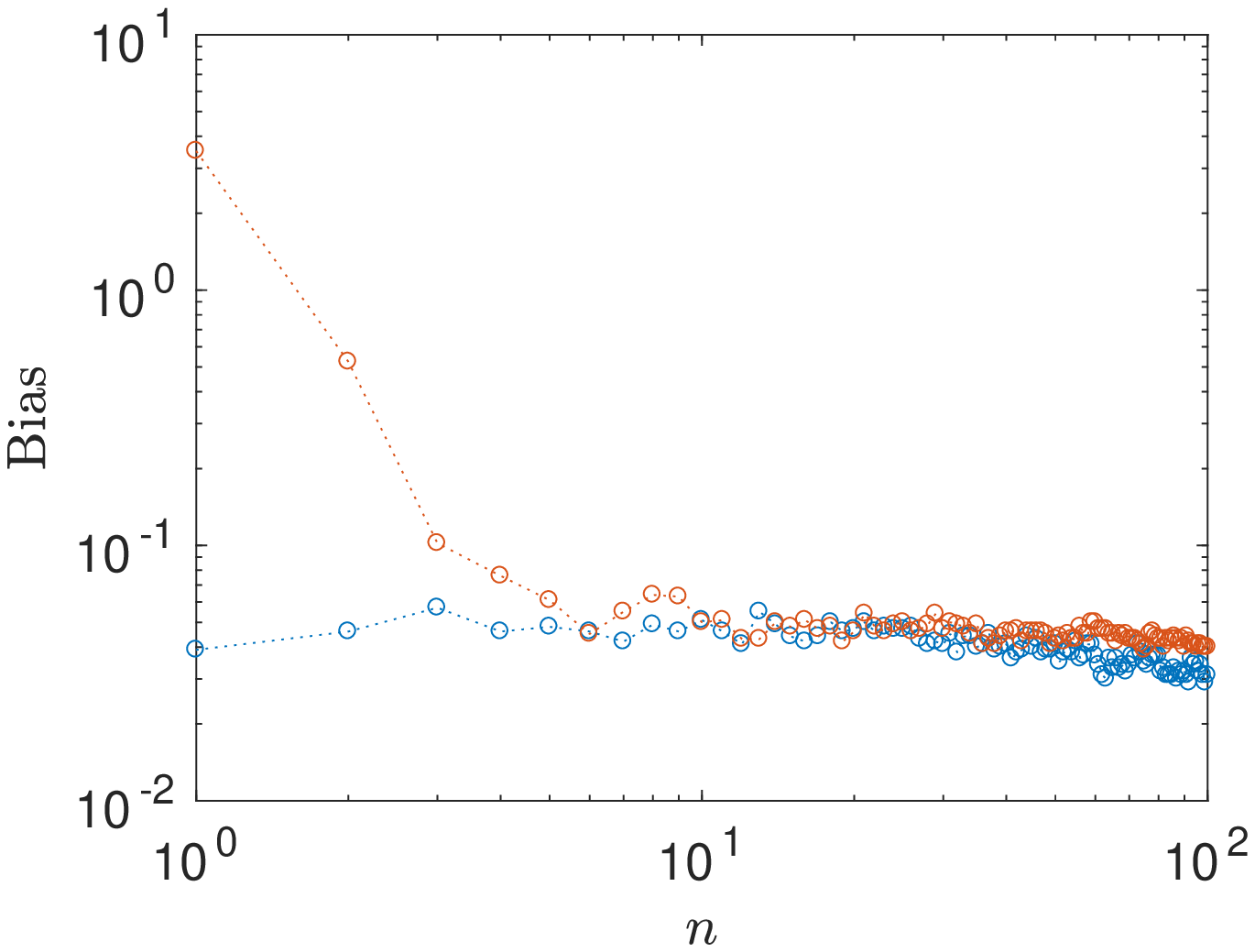}
  \includegraphics[width = 0.49\textwidth]{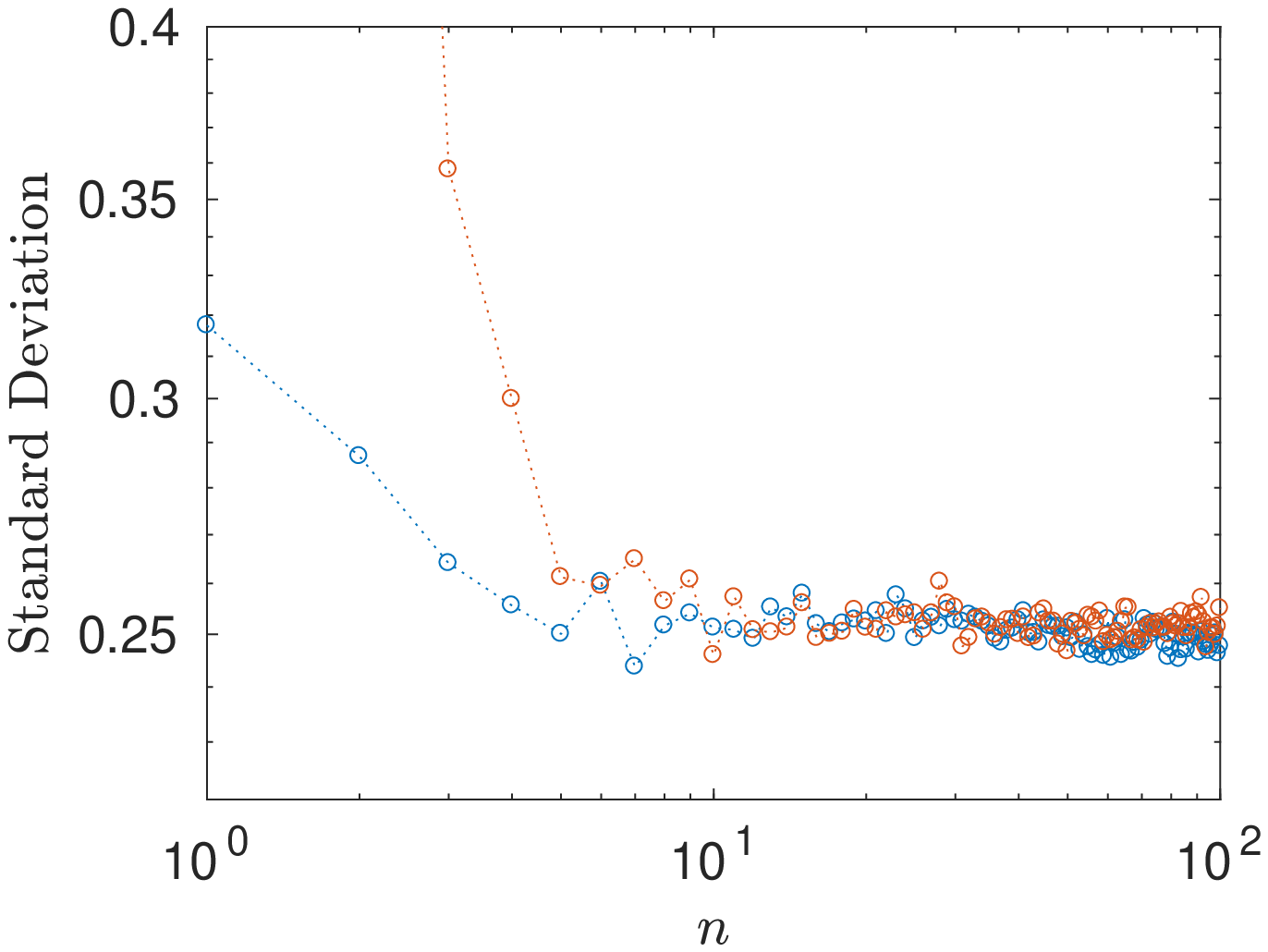}
  \caption{Sample statistics of $\hat{\alpha}$ as the number of observations, $n$, varies. Blue: our policy, and red: observations spaced uniformly in time.}\label{fig:n}
\end{figure}
\end{enumerate}

\subsection{Algae and Rotifers}

The next diffusion we study is motivated by an ecosystem. Algae and miniscule rotifers that feed off the algae are trapped in a static chemical environment called a \emph{chemostat}. If we assume that the rotifers are satiable---meaning that their rate of consumption does not increase without bound in the presence of more and more algae---then the continuum limit of the two species' populations is commonly modeled with a \emph{Rosenzweig-MacArthur} system. We add Brownian noise to simulate stochastic effects due to finite populations. In doing so, we obtain
\begin{gather}\label{eq:rma}
\begin{aligned}
  dx &= x\big( r -x - \frac{gy}{K + x}\big) dt + \sigma dw_1 \\
  dy &= y\big( \frac{gx}{K + x} - d\big) dt + \sigma dw_2.
\end{aligned}
\end{gather}
The variable $x$ (respectively, $y$) represents the density of algae (rotifers) per unit area.

In keeping with \cite{yoshida}, we set the parameter $d = 1$, $g = 2$, and $r = 1.8$. At these values, the noise-free system has an attracting limit cycle that collapses in a Hopf bifurcation as $K$ increases to $0.6$, where $K$ is the algal density when the rotifer kill rate is at its half-maximum. Ecologically this means that the numbers of algae and rotifers will settle into boom-bust cycles, with the cycles becoming smaller as $K$ approaches its critical value. This bifurcation parameter, $K$, is our parameter of interest. While \cite{yoshida} uses a value of $0.3$, we take $K = 0.5$ to reduce these cycles to a period of approximately nine days.

We take $\sigma = 4 \times 10^{-4}$ and start the system from $\textbf{x}_0 = (0.4, 0.2)$ so that the algal population is twice as large as the number of rotifers. We observe the chemostat biweekly for four weeks (i.e., $n = 8$ and $\tau = 28$). 

\begin{figure}
  \centering
  \includegraphics[width = \textwidth]{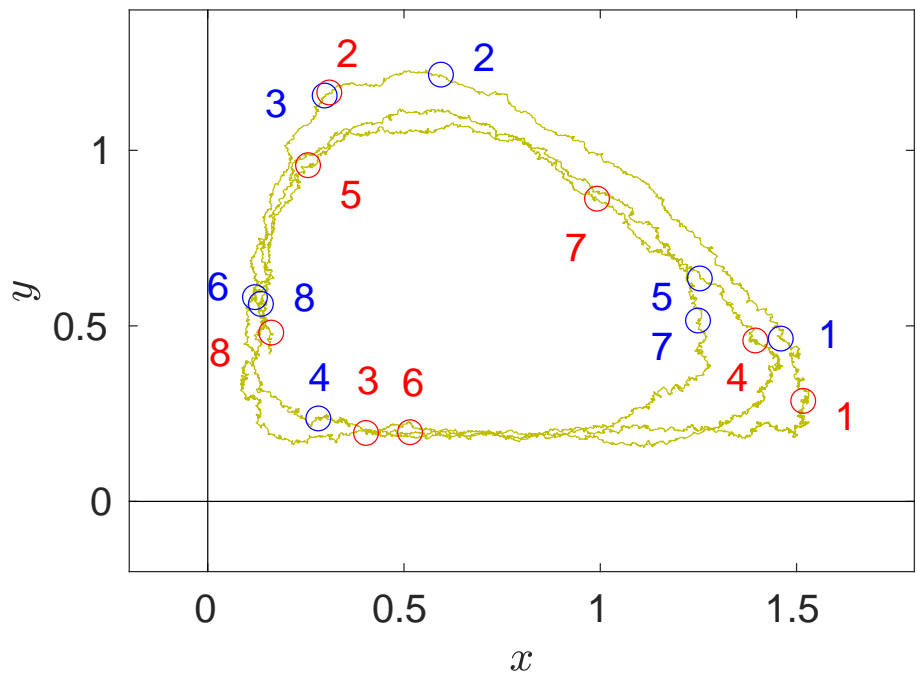}
  \caption{A sample path of the Rosenzweig-MacArthur system \cref{eq:rma}. Our policy's observations and those spaced uniformly in time are circled in blue and red, respectively. The two sets of $n = 8$ observations are labeled from first to last.}\label{fig:rma}
\end{figure}

The analogs of \cref{fig:pkm} and \cref{tbl:pkm} are presented as \cref{fig:rma} and \cref{tbl:rma}.\footnote{$S = \texttt{\{-0.2:0.025:1.8\}} \times \texttt{\{-0.2:0.025:1.4\}}$, $\delta = 10^{-5}$, $\gamma = 2800$, and we use the grid of candidate values $\Phi = \texttt{0.25:0.01:1.25}$.}
\begin{table}
  \caption{Sample statistics of $\hat{K}$ across 100 actualizations with $n = 8$.}\label{tbl:rma}
  \centering
  \begin{tabular}{ c || c | c }
    & Bias & Standard Deviation \\ \hline\hline
	Policy           & 0.0029 & 0.0148 \\
	Uniformly Spaced & 0.0030 & 0.0181
  \end{tabular}
\end{table}
As in the previous example, our policy improves estimates of the parameter of interest; however the gains here are less drastic. The biases of the two sets of estimates in \cref{tbl:rma} are essentially the same, but our policy reduces the standard deviation of estimates by about 18\%. Overall this translates to an MSE that is approximately 32\% smaller.

To develop a better intuition for how the policy behaves on a perturbed limit cycle, we turn to a different system with a more interpretable geometry.

\subsection{A Slow-Fast System}

We consider the stochastically-perturbed Lienard system \cite{strogatz}
\begin{align}
  \epsilon dx &= (y + x - \frac{x^3}{3}\big) dt + \sigma dw_1 \label{eq:vdp1} \\  
  dy &= -x dt + \sigma dw_2.\label{eq:vdp2}
\end{align}
Our parameter of interest is $\epsilon = 0.05$ and we set $\sigma = 0.1$. 

\begin{figure}
  \centering
  \includegraphics[width = \textwidth]{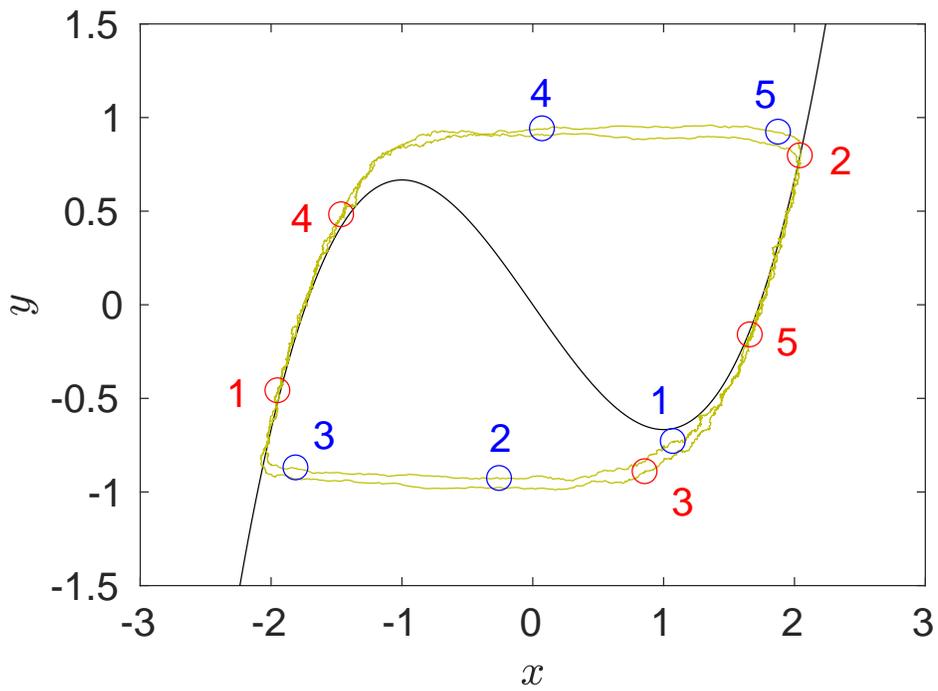}
  \caption{A sample path of the Lienard system \cref{eq:vdp1}--\cref{eq:vdp2}. As before, the observations chosen by our policy and those spaced uniformly in time are circled in blue and red, respectively. The two sets of $n = 5$ observations are labeled from first to last.}\label{fig:vdp}
\end{figure}

Differentiating the top equation and rescaling time reveals that \cref{eq:vdp1}--\cref{eq:vdp2} is equivalent to the \emph{van der Pol oscillator} when $\sigma = 0$:
\begin{equation*}\label{eq:vdpe}
  \ddot{x} + \mu(x^2 - 1)\dot{x} + x = 0.
\end{equation*}
For a small $\epsilon$ like ours and for $\sigma = 0$, the system \cref{eq:vdp1}--\cref{eq:vdp2} is organized around the cubic nullcline $\dot{x} = 0$. This is because $|\dot{x}| \gg |\dot{y}|$ when the right-hand side of \cref{eq:vdp1} is at least the same size as \cref{eq:vdp2}. Thus trajectories rocket horizontally to an outer branch of the nullcline (shown in black in \cref{fig:vdp}). As trajectories approach a $\mathcal{O}(\epsilon^{-1})$ neighborhood of the curve, $\dot{x}$ and $\dot{y}$ become similarly-sized, prompting solutions to follow the $\dot{x} = 0$ nullcline toward the nearest bend. Close to this region, the orientation of the vector field pushes trajectories away from $\dot{x} = 0$, causing $\dot{x}$ to overwhelm $\dot{y}$ and sending solutions rocketing back to the cubic. The process repeats again, and as trajectories circuit around the nullcline, they squeeze onto an attracting limit cycle whose period is approximately 2.4 for our value of $\epsilon$.

To demonstrate our policy, we initialize the noisy system at $\textbf{x}_0 = (1.75, 0)$, allowing it to be observed $n = 5$ times until $\tau = 5$. This corresponds to approximately two cycles. 

\cref{fig:vdp} shows a generic sample path of our system with the two sets of observations: those chosen by our policy and those spaced uniformly in time.\footnote{$S = \{\texttt{-3:0.1:3}\} \times \{\texttt{-1.5:0.1:1.5}\}$, $\delta = 10^{-5}$, and $\gamma = 1000$.}
The geometry of the system cleanly separates the two types. Uniformly-spaced observations tend to fall on the outer branches of $\dot{x} = 0$ since the system spends most of its time there. But notice that $\epsilon$ controls the separation of trajectories from the nullcline and is most strongly felt during the fast transitions between branches. These fleeting, information-rich jumps are precisely where our policy tries to place observations.

\cref{tbl:vdp} shows the sample statistics when using the candidate grid $\Phi$ stretching from $0.01$ to $0.2$ in increments of $0.01$. The bias of estimates from our design are an order of magnitude smaller than those that use uniformly-spaced observations, and the standard deviation is a notable 75\% smaller.

\begin{table}
  \caption{Sample statistics of $\hat{\epsilon}$ across 100 actualizations with $n = 5$. Rows 3 and 4 use uniform priors over $\Phi$ and are explained in detail in \cref{sec:dependence}.}\label{tbl:vdp}
  \centering
  \begin{tabular}{ c || c | c }
    & Bias & Standard Deviation \\ \hline\hline
	Policy           & -0.0002 & 0.0020 \\
	Uniformly Spaced & -0.0022 & 0.0080 \\ \hline
	Averaged Policy  &  0.0017 & 0.0047 \\
	Value Iteration  &  0.0003 & 0.0017
  \end{tabular}
\end{table}
\section{Parameter Dependence}\label{sec:dependence}

The transition matrix, $P$, of the approximating Markov chain depends on $\theta$, and since our policy is a function of $P$, it too depends on $\theta$. (This is to be expected: the best way to observe a diffusion should depend on the diffusion.) However, for our design problem, the $\theta$ dependence is problematic because it requires we know the very parameter we are trying to estimate.

Our goal in the previous section was to demonstrate the proposed policy. To that end, running the policy with $\theta$ set to its true value is wholly appropriate. But for bona fide applications, the $\theta$ dependence needs to be addressed.

There are several possible work-arounds. The most straightforward is to run our policy on a best guess for $\theta$. A more sophisticated strategy is to average our optimization objective over a $\theta$ prior, $\pi$; i.e., take
\begin{align}\label{eq:avg1}
  \hat{t}_i(s, \textbf{x}) = \argmax_{t \in (s, \tau)} \mathbb{E}_{\theta \sim \pi} \left\{ \mathcal{I}(t - s, \textbf{x}) + \mathbb{E}_{\textbf{y} | (t, \textbf{x})} \mathcal{M}_{n - (i - 1)}(t, \textbf{y}) \right\},
\end{align}
redefining the maximal information to go, $\mathcal{M}$, as
\begin{align}\label{eq:avg2}
  \mathcal{M}_i(s, \textbf{x}) = \sup_{t \in (s, \tau)} \mathbb{E}_{\theta \sim \pi} \left\{ \mathcal{I}(t - s, \textbf{x}) + \mathbb{E}_{\textbf{y} | (t, \textbf{x})} \mathcal{M}_{i - 1}(t, \textbf{y}) \right\}.
\end{align}

This change is simple to implement. The resulting sample statistics for a uniform prior are included as row 3 of \cref{tbl:vdp}. Notice that these numbers fall between the corresponding statistics for the two observation types previously considered. Averaging over the uniform prior is roughly 41\% less variable than spacing observations uniformly in time, but about 135\% worse than running the policy on the true value of $\theta$. 

The obvious improvement to this strategy is to update $\pi$ into a posterior as each observation is recorded. However, because of the nestedness of our dynamic program, we would have to recalculate the policy on the remaining $n - k$ observations with the new $\pi$. Doing so is computationally intensive and cannot be done online for most applications. Therefore we consider a third possibilty for handling an unknown $\theta$.

\subsection{The Value Iteration Alternative} 

The \emph{value iteration algorithm} from the theory of Markov decision processes can be adapted to our design problem, and it is more amenable to online posterior updates.

The algorithm is based on an infinite time horizon and optimizes a different objective, the asymptotic average information:
\begin{align}\label{eq:ai}
  \mathcal{O}(\textbf{t}, \textbf{x}_0) = \limsup_{n\to\infty}\frac{1}{n}\mathbb{E} \left[ \sum_{k = 0}^n \mathcal{I}(t_{k + 1} - t_k, \textbf{x}_k) \right].
\end{align}
Here $\textbf{t} := (t_1, t_2, \ldots, t_n)$ and the expectation runs over all observations $\textbf{x}$ other than the initial condition $\textbf{x}_0$. 

The authors of \cite{leifur} find that the value iteration algorithm, modified for their design problem, runs quickly enough to update $\pi$ online. However, our paradigm of observation times poses greater computational burdens because, in our setting, the algorithm requires repeatedly multiplying $P$ with itself. These multiplications are expensive and are typically too numerous to be precomputed and stored. 

The value iteration algorithm for our design problem is faster than the dynamic program we have proposed. However, whether or not the former evaluates quickly enough for online updates will depend on the application: the cardinality of $S$ will determine the complexity of the bottlenecking operation, and the acceptable run time is decided by the time scale of the diffusion (if it is years, then there should be plenty of time to let code run). We judge that the run time of our value iteration algorithm is too long to conclude that online updates are possible for ``most" applications. Row 4 of \cref{tbl:vdp} shows that value iteration outperforms \cref{eq:avg1}--\cref{eq:avg2} for our slow-fast system. Indeed, it is comparable to the optimal policy we propose, but we stop short of calling it ``better" because the improvement in standard deviation is of the same order as our numerical discretization of the system.

A detailed description of the value iteration algorithm for our design setting is given in \cref{sec:via}.
\section{Conclusion}\label{sec:conclusion}

We have optimally estimated a one-dimensional parameter of an It\^o diffusion using a novel and practical design. In particular, we assume a sample path of the diffusion can be observed adaptively, but only $n$ times
over a finite interval $(0, \tau)$.

This problem is important for two reasons: First, diffusions are commonly used to model a variety of phenomena. Second, in our modern information age, the data-driven specification of model parameters is exceedingly topical. Moreover, our design problem arises naturally for systems that are costly to observe.

To choose the $n$ optimal observation times, we adapt the framework of \cite{hlr} and \cite{leifur}, maximizing the observations' expected Fisher information. We do so with a dynamic program, which we implement numerically by discretizing the diffusion with a locally-consistent Markov chain. (The discretization tacitly assumes that the diffusion is bounded on $(0,\tau)$, but this assumption is not restrictive.) Analogous to \cite{hlr} and \cite{leifur}, the solution to the maximization problem is the policy we propose for choosing observation times.

Numerical simulations suggest that our policy behaves intuitively. It tries to take observations as sample paths travel through regions of state space carrying large amounts of information about the parameter $\theta$. Away from stationarity, our policy can reduce the estimates' variance and bias significantly.

As described in \cref{sec:dependence}, a drawback of our policy is its dependence on the parameter it is designed to estimate. Possible remedies include using a best guess for $\theta$, or a prior over the set of candidate values. We have also discussed an alternate, value-iteration policy that improves the computational tractability of updating the prior online.

In future work, we expect that it will be possible to determine optimal observation times for \emph{partially-observed} diffusions by extending the groundwork of \cite{leifur} and \cite{hlr}. As mentioned in \cite{hlr}, this extension also provides another way to mitigate the aforementioned parameter dependence by treating the unknown parameter as an additional state variable.

A more challenging extension is to develop a policy that is computationally tractable at higher state and parameter dimensions. Currently the computation of our policy is bottlenecked by multiplying $P$, the transition matrix of the approximating chain. These multiplications scale with the cube of the state space dimension. For references, see \cite{powell} and \cite{review}, which have been cited by \cite{hlr}.

\section*{Acknowledgements} This work was partially supported by NSF Grants DMS 1053252, DEB 1353039, and DMS 1712554.

\bibliographystyle{plain}
\bibliography{bibliography}

\appendix
\section{Value Iteration Details}\label{sec:via}

Blackwell optimality ensures that the policy maximizing \cref{eq:ai} also maximizes
\begin{align}\label{eq:dc}
  \mathcal{O}(\textbf{x}_0, \textbf{t}) = \mathbb{E} \left[ \sum_{k = 0}^{\infty} \lambda^k \mathcal{I}(t_{k+1}-t_k, \textbf{x}_k) \right]
\end{align}
if $\lambda \in (0, 1)$ is sufficiently close to one. In words, \cref{eq:dc} is prioritizing the present (i.e., time zero) by discounting each successive observation by an additional factor of $\lambda$. Like \cref{eq:ai}, the expectation is taken over all observations $\textbf{x}$ after the intial condition.

Because the time horizon in \cref{eq:dc} is infinite, its maximizing policy is expected to inherit the Markov property and time-homogeneity from \cref{eq:d}. For each observation pair, $(t_k, \textbf{x}_k)$, chosen by such a policy, the value of \cref{eq:dc}---denoted by $\hat{v}$---satisfies
\begin{align}\label{eq:vi}
  \hat{v}(\textbf{x}_k) = \sup_{\Delta_{k + 1} > 0} \mathbb{E}_{\textbf{x}_{k + 1} | (\Delta_{k + 1}, \textbf{x}_k)} \left[ \mathcal{I}(\Delta_{k + 1}, \textbf{x}_k) + \lambda \hat{v}(\textbf{x}_{k + 1}) \right],
\end{align}
with $\Delta_{k + 1} := t_{k + 1} - t_k$. The value $\hat{v}$ is a function of the initial condition of the process (here $\textbf{x}_k$), but not of $\textbf{t}$, because the observation times are chosen by the policy.

Expression \cref{eq:vi} can be used to compute the optimizing policy and to show that it is unique by interpreting $\hat{v}$ as an element of some function space $\mathcal{F}$. From this perspective, we have a map $f:\mathcal{F}\to\mathcal{F}$; i.e.,
\begin{align*}
  f\big( v(\textbf{x})\big) =: w(\textbf{x}) = \sup_{\Delta > 0} \left\{ \mathcal{I}(\Delta, \textbf{x}) + \lambda \mathbb{E}_{\textbf{y} | (\Delta, \textbf{x})} v(\textbf{y}) \right\}.
\end{align*}
Since $\lambda \in (0, 1)$, $f$ will be a contraction, and so its repeated composition will converge to the optimal policy by the Banach fixed point theorem.

This iteration of $f$ on the value function is the aptly-named value iteration algorithm. Details are spelled out in lines 4 through 7 of \Cref{alg:viu}, which includes posterior updates to a prior, $\pi$, for $\theta$. This prior influences the policy through the additional, outer expectation in the algorithm's definition of $w$.

\begin{algorithm}
  \caption{Value Iteration with Online Updates}\label{alg:viu}
  \begin{algorithmic}[1]
	\State Initialize $w(\textbf{x}) = 0$.
	\For{$k = 1$ through $n$} 
	\State $v = w + \epsilon$ for some tolerance $\epsilon$. 
	\While{$\|w - v\|_{L_1} > \epsilon$}
	\State Set $v = w$.
	\State For all $\textbf{x}\in \text{domain}(v)$, set 
	\begin{align*}
	w(\textbf{x}) = \sup_{\Delta > 0}\mathbb{E}_{\theta \sim \pi} 
	\left\{ \mathcal{I}(\Delta, \textbf{x}) + 
	\lambda \mathbb{E}_{\textbf{y} | (\Delta, \textbf{x})} v(\textbf{y}) \right\}.
	\end{align*}
	\EndWhile
	\State Suppose the sup defining $w$ is attained, and store the maximizer as $\hat{\Delta}_k(\textbf{x})$.
	\State Take Observation $k$ at the optimal time
	\begin{align*}
	t_k = \hat{\Delta}_k(\textbf{x}_{k - 1}) + t_{k - 1}
	\quad \text{with} \quad t_0 := 0,
	\end{align*}
	and store the outcome as $\textbf{x}_k$.
	\State Update $\pi$ with Bayes's rule:
	\begin{align*}
	  \pi_{k + 1}(\theta) \propto p_{\theta}(\textbf{x}_k, \hat{\Delta}_k(\textbf{x}_{k - 1}) | \textbf{x}_{k - 1}) \times \pi_k(\theta).
	\end{align*}
    \EndFor	
  \end{algorithmic}
\end{algorithm}

Of course, it is not numerically possible to take the supremum defining $w$ over a truly infinite time horizon. Thus, in our numerical implementation, we let the supremum run from zero to $\tau/n$ to ensure that the budget of $n$ observations will be spent before the time horizon $\tau$.

\end{document}